\documentclass{article}

\usepackage{amssymb}

\usepackage{epsfig}

\usepackage{lscape}

\usepackage{graphicx,psfrag,amsmath}

\usepackage{yfonts}

\begin{document}

\def\beq#1\eeq{\begin{equation}#1\end{equation}}
\def\beql#1#2\eeql{\begin{equation}\label{#1}#2\end{equation}}

\def\bea#1\eea{\begin{eqnarray}#1\end{eqnarray}}
\def\beal#1#2\eeal{\begin{eqnarray}\label{#1}#2\end{eqnarray}}

\newcommand{\Z}{{\mathbb Z}}
\newcommand{\N}{{\mathbb N}}
\newcommand{\C}{{\mathbb C}}
\newcommand{\Cs}{{\mathbb C}^{*}}
\newcommand{\R}{{\mathbb R}}
\newcommand{\intT}{\int_{[-\pi,\pi]^2}dt_1dt_2}
\newcommand{\Cc}{{\mathcal C}}
\newcommand{\cI}{{\mathcal I}}
\newcommand{\cN}{{\mathcal N}}
\newcommand{\cE}{{\mathcal E}}
\newcommand{\Ca}{{\mathcal A}}
\newcommand{\xdT}{\dot{{\bf x}}^T}
\newcommand{\bDe}{{\bf \Delta}}

\def\ket#1{\left| #1\right\rangle }
\def\bra#1{\left\langle #1\right| }
\def\braket#1#2{\left\langle #1\vphantom{#2}
  \right. \kern-2.5pt\left| #2\vphantom{#1}\right\rangle }
\newcommand{\gme}[3]{\bra{#1}#3\ket{#2}}
\newcommand{\ome}[2]{\gme{#1}{#2}{\mathcal{O}}}
\newcommand{\spr}[2]{\braket{#1}{#2}}
\newcommand{\eq}[1]{Eq\,\ref{#1}}
\newcommand{\xp}[1]{e^{#1}}

\def\limfunc#1{\mathop{\rm #1}}
\def\Tr{\limfunc{Tr}}

\def\dr{detector }
\def\drn{detector}
\def\dtn{detection }
\def\dtnn{detection}

\def\pho{photon }
\def\phon{photon}
\def\phos{photons }
\def\phosn{photons}
\def\mmt{measurement }
\def\an{amplitude}
\def\a{amplitude }
\def\co{coherence }
\def\con{coherence}

\def\st{state }
\def\stn{state}
\def\sts{states }
\def\stsn{states}

\def\cow{"collapse of the wavefunction"}
\def\de{decoherence }
\def\den{decoherence}
\def\dm{density matrix }
\def\dmn{density matrix}

\newcommand{\mop}{\cal O }
\newcommand{\dt}{{d\over dt}}
\def\qm{quantum mechanics }
\def\qms{quantum mechanics }
\def\qml{quantum mechanical }

\def\qmn{quantum mechanics}
\def\mmtn{measurement}
\def\pow{preparation of the wavefunction }

\def\me{ L.~Stodolsky }
\def\T{temperature }
\def\Tn{temperature}
\def\t{time }
\def\tn{time}
\def\wfs{wavefunctions }
\def\wf{wavefunction }
\def\wfn{wavefunction} 
\def\wfsn{wavefunctions}
\def\wvp{wavepacket }
\def\pa{probability amplitude } 
\def\sy{system } 
\def\sys{systems }
\def\syn{system} 
\def\sysn{systems} 
\def\ha{hamiltonian }
\def\han{hamiltonian}
\def\rh{$\rho$ }
\def\rhn{$\rho$}
\def\op{$\cal O$ }
\def\opn{$\cal O$}
\def\yy{energy }
\def\yyn{energy}
\def\yys{energies }
\def\yysn{energies}
\def\pz{$\bf P$ }
\def\pzn{$\bf P$}
\def\pl{particle }
\def\pls{particles }
\def\pln{particle}
\def\plsn{particles}

\def\plz{polarization  }
\def\plzs{polarizations }
\def\plzn{polarization}
\def\plzsn{polarizations}

\def\sctg{scattering }
\def\sctgn{scattering}

\def\prob{probability }
\def\probn{probability}

\def\om{\omega} 

\def\hf{\tfrac{1}{2}}
\def\hft{\tiny \frac{1}{2}}

\def\zz{neutrino }
\def\zzn{neutrino}
\def\zzs{neutrinos }
\def\zzsn{neutrinos}

\def\zn{neutron }
\def\znn{neutron}
\def\zns{neutrons }
\def\znsn{neutrons}

\def\hf{\tfrac{1}{2}}

\def\csss{cross section }
\def\csssn{cross section}

\def\epp{elementary particle physics }
\def\eppn{elementary particle physics}

\def\vhe{very high energy }
\def\vhen{very high energy}

\title{ Behavior Of Very High Energy Hadronic\\ Cross Sections }

\author{L. Stodolsky,\\
Max-Planck-Institut f\"ur Physik
(Werner-Heisenberg-Institut)\\
F\"ohringer Ring 6, 80805 M\"unchen, Germany}

\maketitle


\section{Introduction}
 A classic chapter of \epp appears to be drawing to a close,
and  it seems appropriate  at this point
  to describe, in general terms, some of the issues and how they
have been resolved.

The question is the \sctg of elementary hadrons---strongly
interacting \plsn, notably the proton--- at \vhen. The  question of
the \vhe or
asymptotic behavior of
the \csss for strongly interacting \pls  --``hadrons"--is one of 
simplest to pose, but took the longest to answer. Indeed, it has
taken a  {\it surprisingly} long time to answer, and  one of the
points we would like to address is why.

We say {\it `surprisingly'}
because  very high \yy data  for  protons or anti-protons have been
available for decades, at the big accelerators or from cosmic rays.
The \yys available, expressed in terms of the center-of-mass  \yy
$W$   have long been well above any evident scale
one might associate with strong interactions. For such a scale, one
might consider the 
proton mass itself, about
1 GeV. Or perhaps the $\Lambda$ parameter of QCD,
the field theory  presumed to underly  strong interactions. But
this is even less, only
around $0.2$ GeV. Otherwise it's hard to think of any obvious \yy
scales.  On the other hand, the  Fermilab TeVatron 
surpassed  $W=1\, TeV= 1,000 \,GeV$
years ago, and CERN's LHC recently reached $W=13\, TeV =13,000
\,GeV.$
By any measure, one would have thought, a simple pattern or picture
should    certainly  have
emerged long ago.  Even given that the quantities in question  vary
only slowly, logarithmically, this  great discrepancy is a real
connundrum. Is there perhaps a new, higher mass scale waiting  to
be discovered in
the wings?  Those patient
enough  to read to the end  will find there is an answer to this
puzzle, but from an
unexpected direction.

Our  main source of  information on \vhe
 comes from proton-proton (p-p) or 
antiproton-proton ($\bar{p}$-p) studies at accelerators, or from
 cosmic rays. 
While many studies of hadron interactions exist,   we will
concentrate on these channels, but it is
probable that our main points would apply equally well to others,
such as $(\pi$-p) or (K-p), and with suitable account of
`vector dominance',
to $(\gamma$-p) reactions.

In all these channels there is a low \yy region,  where the \csssn
s are characterized by various resonances and particularities of
the individual channel.
 But for   center-of -mass  \yys  $W$ above the tens of GeV, a
smooth behavior for the \csss sets in, with probably a
universal behavior in all channels.

 What is this general behavior and how is it to be described...and
perhaps to be explained? 
Fig 1  shows a plot of \vhe p-p and  $\bar{p}$-p data, as a
function of center-of-mass \yy W, for the  total (upper curve),
the inelastic (middle curve) \csssn s  and (lower curve) their
difference,  the elastic \csssn.  Aside from the evident fact that
the \csssn s  are
increasing with  \yyn, the plot is rather featureless and there
seems
to be no particular relationship between the three curves.
 However, as  we shall see, there is nevertheless  a
hidden simplicity which describes the data.

\begin{figure}[h]
\includegraphics[width=\linewidth]{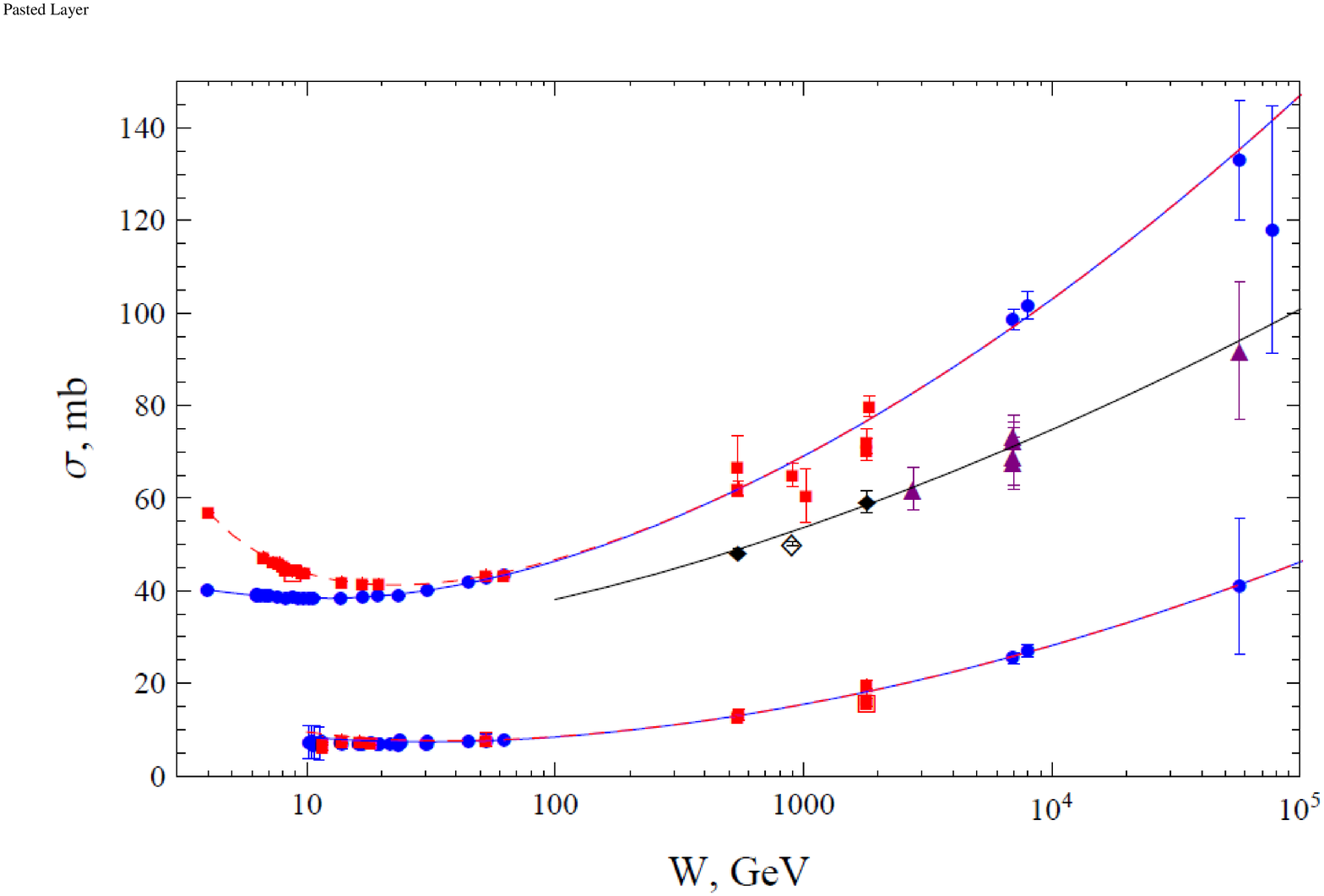}
\caption{ Behavior of proton (blue, purple), antiproton
(red,black)-proton \sctg up to \vhen,
shown as a function of the center-of-mass \yy $W$.  The curves  are
for the total (upper), inelastic(middle), and elastic (lower) 
\csssn s. From ref \cite{comp}.}
\label{css}
\end{figure}

\section{Rise of the Cross Sections}
The  increase of the \csssn s with \yy is in itself not a trivial
point, and 
until the early 70's, when the rise was first seen at CERN's ISR,
the
opinon was  often heard 
that the \csssn s 
would approach a constant limit at \vhen. One might think, after
all, that an incoming proton, regardless of its \yyn, is just
hitting a fixed, unchanging  target, an object of constant size.
However this
turned out not to be the case, as one sees. 
 The
question is certainly an old one, with proposals going back at
least to Heisenberg in the 50's  \cite{heis}, who suggested that
the total \csss $\sigma$ should
increase as the square of a logarithm,
\beql{ln}
\sigma \sim ln^2 (W/W_o) \, ,
\eeql
where $W_o$ is some \yy scale parameter.

  The idea underlying \eq{ln} is a field 
theoretic one, and its validity can be interpreted as yet
another manifestation of field theory in fundamental interactions.
 The argument is essentially that with greater \yyn,
a proton can excite a  target,  (e.g. produce a meson off 
  another proton) from ever greater distances. Since the
radius of interaction $R$ is thus growing, so is the \csssn, which
goes as $R^2$. Thus \eq{ln} corresponds to a logarithmic increase
in the range of interaction.

 This behavior is actually not surprising and  has been known
 for a long time in the more ordinary
process of the ionization produced by a charged \pl passing 
through matter. There is a phenomenon  called ``the relativistic
rise"\cite{rise} where due to the relativistic boost of 
the electric/magnetic fields  around a very fast charged \pln, 
it can, as its \yy  goes up, eject electrons or excite atoms
further and further away from the charge.

 Analogously, with increasing  \yy one can expect an
increasing intensity or \yy density in the fields surrounding the
highly relativistic proton and
an increasing  probability of  ``ejecting''  something  or exciting
a target at ever greater distances. Thus there is an
increasing radius of interaction.

 However, in hadronic interactions like proton-proton \sctg there
is an important difference vis-a-vis the ``relativistic rise'' in
ionization. There, one is concerned with the relativistic `boost'
of the long--one could say infinite--range coulomb field. Here, the
field around the proton 
is  of short range or  Yukawa-like:  $\sim
(1/r)e^{-\mu r}$. This introduces $\mu$, a mass or inverse
length parameter (We use natural units $\hbar=c=1$ where a mass is
also an inverse length).

 This exponential cutoff means that even if the fields  boost as
some power $p$ of the \yyn, it will take a high \yy for them to
obtain a significant value at large distances.  If some threshold
value is required to have significant  \pl production at a distance
$ r_{max}$ we will have the leading condition
\beql{thld}
 W^p \times e^{-\mu r_{max}}\geq threshold \,,
\eeql
a relationship connecting  the \yy and the effective maximum
range of interaction $r_{max}$. Taking the log  we
obtain $r_{max}=constant \times ln W$.   Since the \csss
$\sigma$ goes as $r_{max}^2$, one has \eq{ln} .

 A detailed concretization of this general argument is provided 
 by the work
of the Apsen group \cite{comp}, where the field density around
the proton is taken
from electromagnetic form factor measurements  and an eikonal
methods is used \cite{chouyang}. With
a power increase for the interaction,
the asymptotic $ln ^2W$
behavior is indeed obtained. Furthermore, good fits are found to
the
other quantities such as the shape  and \yy dependence of the
elastic diffraction peak induced by the absorbtion \cite{comp}.

 Interestingly, more formal arguments,  based on the analytic
properties of \sctg amplitudes in the Mandelstam representation,
led to the conclusion that
\eq{ln} represents in fact the fastest growth possible \cite{fm}
consistent with analyticity.  Further steps along these lines even
led  to a an upper bound on the coefficient of the $ln^2W$, with a
dimensional coefficient characterized by the pion mass 
$\pi/m_{\pi}^2\approx 60\, mb$.  
 It should be stressed that these mathematical results
represent upper bounds and that there is nothing  {\it a priori}
wrong with a smaller coefficient or a slower behavior.

For many years, the question of the experimental validity or not of
\eq{ln} was unclear. This is due to the fact that the logarithm is
a very slowly changing function, so that even the great efforts in
achieving higher $W$'s over the decades led to only relatively
modest changes in the log's. Furthermore,
whatever the final \vhe behavior, 
there are non-leading terms which will only slowly disappear,
making the extraction of the leading behavior non-trivial.  Thus
despite the enormous developments in
accelerator technologies, it was not easy to extract and
distinguish one model for the asymptotic \csssn s from another.  

\section{The Growing "Black Disc''}
Over the years, different groups developed fits to the high \yy
data \cite{groups} . Finally some 
 clarity in the situation began to emerge in the last decade 
when M.M. Block and Francis  Halzen of the Aspen group noticed
\cite{BH} something very interesting about their fits. Not only
could one get good fits with a leading  $ln^2W$ term, but also
the coefficients  of the  elastic and total \csssn s  for these
terms were accurately in the ratio 1:2.

The fact that the ratio
$\sigma(elastic)/\sigma(total)$  approaches one-half 
\beql{lim}
\frac {\sigma(elastic)}{\sigma(total)}\to
\frac{1}{2}~~~~~~~~~~~~~~~~~~~~~~~~~~~~~~~W\to \infty
\eeql
corresponds to the standard `black disc' limit, enshrined in 
classical optics as ``Babinet's principle''.  One has
a  disc which is totally `black',that is, everything hitting it
 is completely
absorbed, giving the inelastic \csssn. At the same time this
`absorption' creates a `hole' in
the incoming  wave front, leading to an elastic \sctg which  has
the  same
\csss \cite{bab}. Thus one has the situation
$\sigma(total)=\sigma(elastic)+\sigma(inelastic)=2\sigma(elastic)$,
and so \eq{lim}.

From the fits it thus appears that in the \vhe limit of proton-
proton \sctg one approaches a text-book
`black disc' and with an \yy dependence in accordance with \eq{ln}
and  the  Froissart bound.

\section{Multiplicity and Cross Sections}
 It would be good to have some further physical support of the
general picture. If, as
argued above, the growth of the \csss with \yy is connected
to the possibility of producing \pls at ever-increasing distances
or impact parameters, then there should be some connection between
the growth of the \csss and the number of \pls  produced in a
collision.  This
suggests looking at the multiplicity $N(W)$, the average number of
\pls produced in a collision, which is also increasing with \yyn.
  A proposal along these lines \cite{leo} was
that at
\vhe the two quantities  should  grow in parallel:
\beql{one}
\sigma \propto N~~~~~~~~~~~~~~~~~~~~~~~~~~~W \to \infty.
\eeql
 
\begin{figure}[h]
\includegraphics[width=\linewidth]{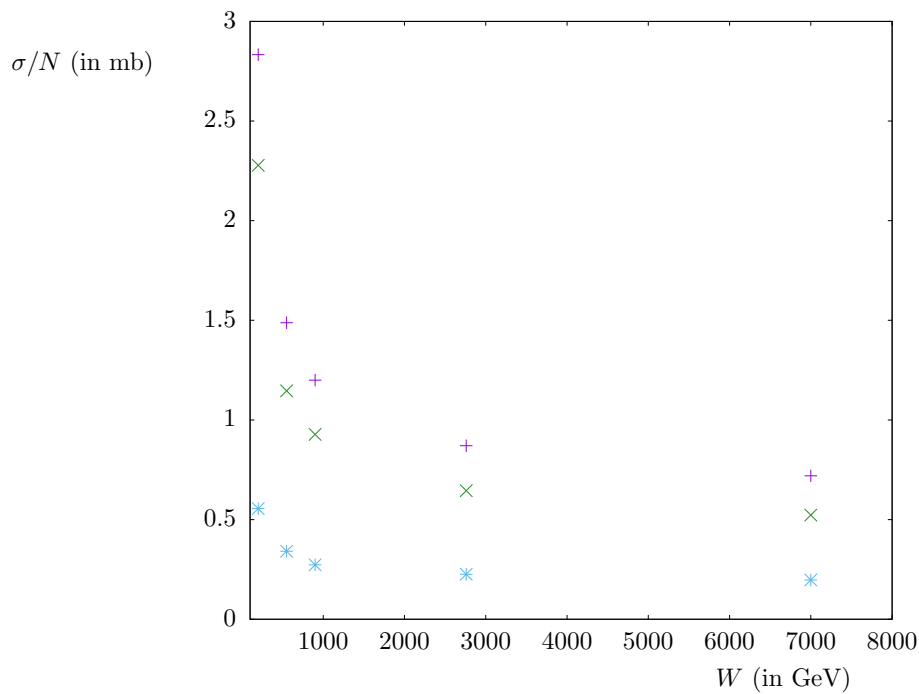}
\caption{ Ratios of the total (red +), inelastic (green x) and
 elastic (blue *) \csssn s to 
the total
pion multiplicity (which is approximately the total multiplicity),
in mb vs. the center-of-mass \yy $W$ in GeV. From \cite{plb}.}
\label{ratios}
\end{figure}
 
For this to be true, in view of \eq{ln}  the leading behavior for
$N$  should also be as $ln^2W$ at \vhen. In fact, a plausible fit
where this is the case is possible using LHC and lower \yy \pl
production data
\cite{plb}. Dividing $\sigma$ by $N$ one then has a certain
constant \csss per
produced \pl at \vhen . The  fit leads to 
$0.31\,  mb$ of total \csss  per  pion.
 Like  the \csssn, the fit for $N$ has the feature
that along with the  $ln^2W$ there is non-leading $ln W$ term with
a large negative coefficient \footnote{Technical note: Because of
the identity $ln^2(W/W_o)=\bigl(ln(W/W'_o)-
ln(W_o/W'_o)\bigr)^2=ln^2(W/W'_o)-2ln(W/W'_o)ln(W_o/W'_o)
+ln^2(W_o/W'_o)$, one can trade a linear ln term for a change
in the scale in the argument of the  $ln^2$ term. Our
statements about a large
non-leading term are relative to the use of a moderate scale  in 
the $ln^2$ terms, which is $1\, GeV$ in the fits we quote. Note
this
ambiguity of representation has no effects on the {\it coefficient}
of the $ln^2$ term.} Since at presently available \yys 
``Asymptopia'' is
still far,  the $\sigma/N$ are not all
constant and we do not yet have the limiting 
$\sigma(elastic)/N=\hf
\sigma(total)/N$.
Interestingly, however, $\sigma(elastic)/N$ has reached its
 limiting value and appears to be constant. These
features are  shown in  Fig.\,\ref{ratios}. There seems to be
experimental support for  \eq{one} and for the idea of a  simple
connection between the rise in the \csssn s and rise of the
multiplicity.

\section{The `Edge' of the Proton}
We thus arrive at a simple picture: At \vhe  the proton looks like
a simple `black disc' where the elastic \csss $ \sigma(elastic)$ is
half that of the total \csss $\sigma(total)$,  as somebody who had
\sctg theory in their \qm 1 course might
have guessed. And, as he or she might not have guessed, if they
didn't know about \eq{ln}, the radius of this
disc is growing logarithmically.

\begin{figure}[h]
\includegraphics[width=\linewidth]{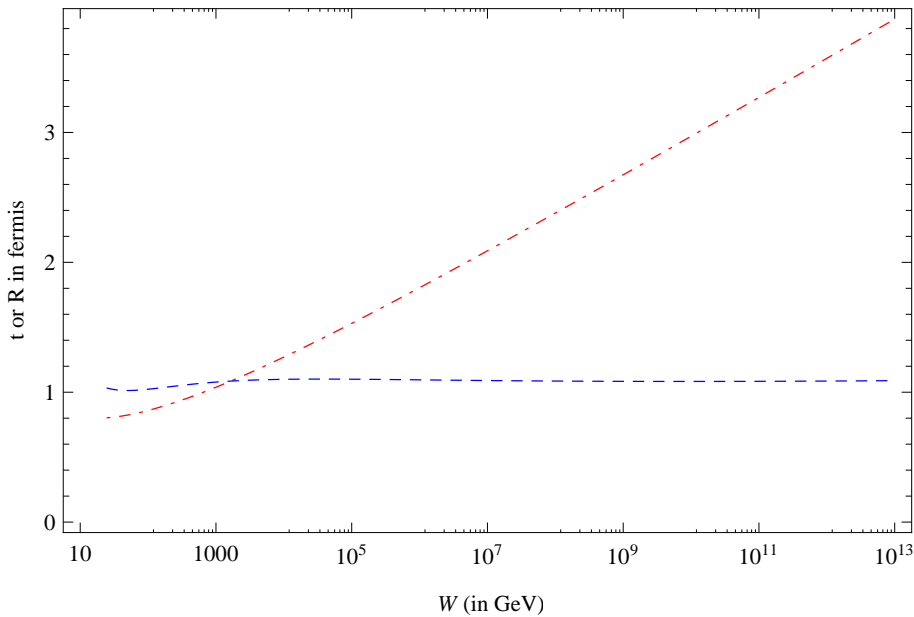}
\caption{The ``edge'' and the ``disc''. The dashed (blue) line is
a plot of the ratio \eq{csssc}, 
 representing  $t$, the effective thickness  of the edge.  Its
constancy  exhibits the 
\yy independence of the edge.  For
comparison the dashed-dotted line (red) represents  the black-disc
radius $R$ inferred from the total \csssn, 
$R=\sqrt{\sigma(total)/2\pi}$. The units are  in fermi=fm $=10^{-
13}cm$.  From ref \cite{edge}. 
}
\label{ratio}
\end{figure}

 But along  with this pleasant
picture come two questions:

A) Why did it take so long for this picture to emerge, until \yys
in the many TeV range? As said above, these \yys are a thousand or
ten thousand times greater than any obvious mass scale. 

B) A simple "black disc'' with an abrupt hard edge seems rather a 
mathematical idealization. Wouldn't it be more physical and
realistic to have some kind of a  soft edge, with a gradual
transition from total opacity for  central collisions to
complete transparancy at large distances?

 It turns out   the answer to A) comes from examining  B).

To examine B) we need  some quantity which will isolate the    
possible `edge' hidden in the experimental information. This
can be done as follows \cite{edge}. Both of the \csssn s
$\sigma(total)$ and $\sigma(elastic)$  may be written as a
sum of contributions over impact parameter $b$.  We consider
the quantity  $(\sigma(total)-2\sigma(elastic))$, which would be
zero at all $b$ for the idealized  "black disc''. In this impact
parameter representation, this difference  peaks in the  vicinity 
of the radius \cite{edge}. This
occurs because, with only a small real part to the amplitude, 
both  $ \sigma(total)$ and $\sigma(elastic)$ are given by the same
amplitude, the first linearly via the optical theorem, and the
second quadratically via squaring the amplitude.  In terms of 
a transparency $\eta(b)$ the difference 
$\sigma(total)-2\sigma(elastic))$
is  given by an integral over impact parameter $b$, namely
$4\pi\int_0^\infty \eta(b) (1-\eta(b))\,b\,db $. It will be seen
that as $\eta(b)$ varies from near zero at $b=0$ to one at large
$b$, the integrand  goes from zero to zero  with a peak in
the middle, near where the idealized  edge would be. This makes it
a suitable quantity for isolating an ``edge''. 

 Furthermore if we normalize to
the radius defined by the total
 \csssn, $R=\sqrt{\sigma(total)/2\pi}$, one finds that
the  quantity
 \beql{csssc} 
t=\frac{\sigma(total)-2\sigma(elastic)}{\sqrt{(\pi/2)\sigma(total
)}}
\eeql
represents the `thickness' of the `edge'\cite{edge}.

One notes that this quantity is nicely   constructed from
experimental quantities only. Thus it is indepedent of the fitting
procedure, as long as the fits go through the data and for \yys
where data exists,
and is independent of any theoretical prejudices.

One may use the fits  of Fig \,\ref{css} to
evaluate \eq{csssc}. The result is shown in
Fig\,\ref{ratio}. The blue dashed line is the value of $t$.
 For comparison the radius corresponding to
the total \csss $R=\sqrt{ \sigma(total)/2\pi}\,$ is also shown  as
the  red dashed-dot line.

\begin{figure}[h]
\includegraphics[width=\linewidth]{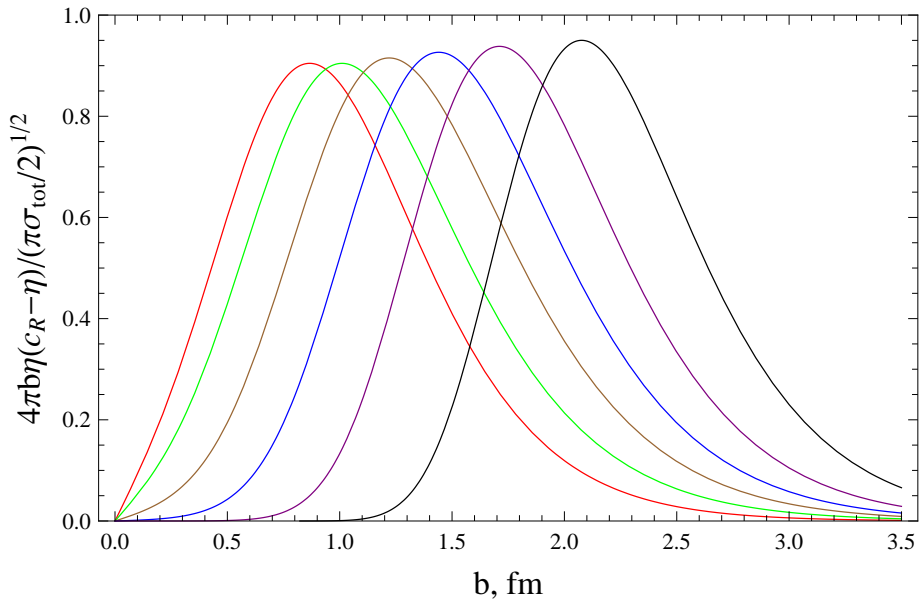}
\caption{  The edge integrand, the quantity whose integral over $b$
gives the ratio \eq{csssc}. It is constructed from the impact
parameter amplitudes which have been fit to give the various
features of elastic \sctg and total \csss data. The plot shows
how the ``edge'' remains approximately constant and  moves out in
$b$ as the black disc expands. The different colors correspond to
\yys $W$ from 30 to $10^8\, GeV$. The quantity $C_R$ deviates
slightly from 1 to account for a small real part of the elastic 
amplitude.  From Fig 9 of ref \cite{edgmove}. 
}
\label{moveplot}
\end{figure}
 One observes that $t$ is constant  with \yyn,
and has the very reasonable value of  $t \approx 1.1\, fm $.
This is the hidden simplicity behind  the seemingly featureless
Fig\,\ref{css}. What  the data actually represent is just a
logaritmically expanding black disc, with a constant  edge.

To see what the ``edge'' looks like, one may also use the further 
information provided by elastic \sctg data at non-zero angles to
reconstruct the 
elastic amplitude as a function of $b$. Fig.\,\ref{moveplot} 
shows the resulting edge integrand, the quantity whose integral
gives the ratio \eq{csssc}. One sees how the ``edge''remains
roughly constant and moves out as the ``disc'' expands. (The
quantity $C_R$ is not exactly 1 to take account of the small real
part of the amplitude.)

\section{Remote Asymptopia}
 We seem to have arrived at a simple and 
satisfying picture. There is a black disc, logarithmically
expanding, in analogy with the ``relativistic rise" of ionization, 
and this disc has a constant, smooth edge.

However, a puzzle remains. These simple features still do not stand
out very clearly, even at the present very high LHC \yysn; the
approach to  
``Asymptopia'' is very slow. In Fig. 1 we are still far from the
limit $\sigma(elastic)=\hf \sigma(total)$ and the same holds for
the $\sigma/N$ ratios in Fig.\,\ref{ratios}.
The limits will finally be reached,  according to the fits,  but 
why is it that
it takes \yys $W$  a thousand or ten thousand times greater than
any evident \yy or mass scale  for this to begin  to happen? (I
also remind the
non-specialist that $W$  is a center-of-mass \yyn, so that it takes
even longer in terms of the laboratory \yyn, which is the relevant
\yy for cosmic rays or fixed-target experiments.)

An explanation seems to be provided by Fig.\,\ref{ratio}.
``Asymptopia'' will be reached
when the  leading ``disc'' is distinctly larger than the 
subdominant ``edge'', and  Fig.\,\ref{ratio} shows this will not
happen until  $W$ is  at least in the multi-TeV
regime. Indeed the complete dominance  of the ``disc'' is delayed
to \yys that  probably will never be reached in accelerator or
cosmic ray experiments. 

\section{The ``Edge'' versus the ``Disc''}
But perhaps should we say ``description" instead of
``explanation''? Fig.\,\ref{ratio} is certainly very interesting,
but one  may rightly say it  just shifts the question.  We would
now like to know  why
 it is  that the ``disc'' is so small or, 
alternatively, the ``edge'' so big?  In terms of \csss parameters
the  $1.1\, fm$ thickness $t$  of the
edge corresponds to $\pi (1.1 f)^2=37\, mb$ while the coefficient
of the $ln^2$ term for $\sigma(total)$ is only the relatively tiny
$1.1\,mb$\cite{newexptl}. 

This brings us to the intriguing question as to the nature of the
``edge''. One might entertain various speculations as to its
origin or makeup. One suggestion is  that $t$ is associated
with the length that the  color string of QCD can be
stretched before it breaks \cite{ros}.

Another very  interesting  possibility  is that the `edge' has to
do with the exchange of a  pion \cite{bj},  the lightest hadron.
Firstly, the 
$t=1.1\, fm $  corresponds well with the compton
wavelength of the pion, $1/m_\pi\approx 1.4 \,fm$. More
significantly, this would put the slow approach to ``Asymptopia" in
a new light and offers an amusing resolution to the puzzle of the
mass scales. The pion is quite light compared to other hadrons.
Indeed, according to 
chiral symmetry, which plays an important role in low \yy
hadron physics,  the pion  should be thought of as  having
initially zero
mass, before acquiring a finite mass  via small
corrections.

From this point of view, we would say that the big ``edge''  comes
from the ``almost zero'' mass of the pion.
  Hence the slow approach to ``Asymptopia'' 
originates not from some hidden high mass scale, but  on the
contrary, in the
existence of a very low scale, which covers up the ultimately
leading behavior until \vhen . Thus while the factor $\pi
/m_{\pi}^2$, long thought to characterize the high \yy 
\csssn \footnote{These considerations also provide an interesting
view on
the question \cite{eduardo} of the Froissart bound in
 the chiral limit $m_{\pi} \to 0$. If the asymptotic $ln^2 W$
really has
as its dimensional prefactor something involving $ 1/m_{\pi}^2$ as
in \cite{fm}
 this would blow up as  $m_{\pi} \to 0$. On the other hand, our
discussion suggests that reducing $m_{\pi}$ gives an increase in
the size of  the ``edge''. Thus a very small $m_{\pi}$, results not
necessarily in a change of the asymptotic term, but rather in the
removal of ``Asymptopia'' to a very 
high \yyn.}, is certainly there, it gives the contribution of the
non-leading edge, and at high enough \yy is finally overtaken
   by the  $(1.1\, mb)\, ln^2W$ of the ``disc''.

 This leaves us with the problem of explaining  the $(1.1\, mb)$.
It seems to have no connection with $m_\pi$, but it does closely 
resemble the scale which one  gets with
the ``usual"  hadrons with masses around a GeV:  $\pi/GeV^2
=1.2\, mb$. 
  It is a challenge to theory
to provide a calculation of the $1.1\, mb$, which  ultimately
gives the  \csss at highest \yys  and so represents a fundamental
parameter of hadron physics.

\vskip1.5cm

This article is dedicated to the memory of Marty Block, who after
a very long and productive career, passed away in July 2016.

\end{document}